# Roche tomography of CVs

**Robert Connon Smith**

*University of Sussex, UK*
*Falmer, Brighton BN1 9QH*
*E-mail:* `r.c.smith@sussex.ac.uk`

Roche tomography is a technique designed for mapping the intensity distribution over the surface of binary star components that fill (or nearly fill) their Roche lobes and so are both rotationally and tidally distorted. It builds on and develops the successful techniques used for mapping rapidly rotating single stars, which have revealed the presence of starspots. It has so far been applied to only a small number of cataclysmic variables, and a few other related systems. This paper reviews the technique and presents maps of spots on most of these systems. The review has been written with the help of contributions of material from Chris Watson and Colin Hill of Queen's University, Belfast.

## 1. Introduction

Roche tomography was first proposed by Rutten & Dhillon [1] as a method of mapping the line intensity distribution over the surface of Roche-lobe-filling components of interacting binaries, and hence discovering starspots. Both emission and absorption lines may be mapped, and the technique takes account of the Roche geometry as well as of limb- and gravity-darkening. For absorption lines, the method uses the changes in the line profile shape with orbital phase as spots cross the disc of the lobe-filling component. The changes arise because the line is rotationally broadened, and each point in the line profile corresponds to a different line-of-sight component of the rotation speed. Any particular spot will produce a 'bump' (Illustration 1) over a small range of wavelength (velocity) within the intrinsic line profile; the bump corresponds to an increase in relative flux in the line, because the local continuum flux is decreased by the presence of the (dark) spot. The velocity range corresponds to the longitude range of the spot. The 'bump' will be seen to travel across the profile as the star rotates during its orbital motion, because its mean line-of-sight velocity changes with observed longitude. A maximum entropy code is used to construct the most likely brightness distribution over the visible surface.

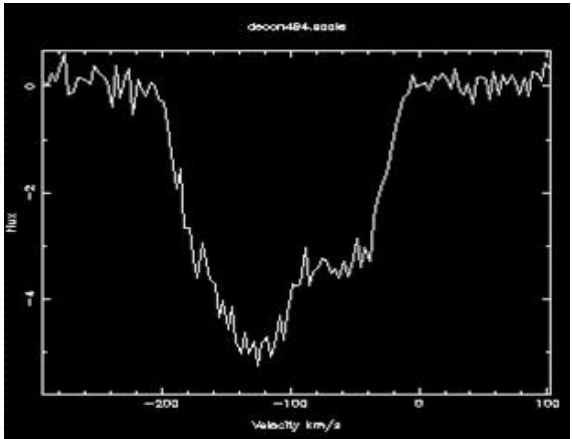

*Illustration 1: A mean line profile for AE Aqr, showing a 'bump'*

The method makes a number of assumptions: the stellar surface coincides with the Roche lobe; the orbit is circular and the lobe-filling star is synchronously rotating; the intrinsic line profile does not change, and arises solely from the lobe-filling star; there are no intrinsic variations during the observations; the binary parameters are known; the data have been slit-loss corrected. With these assumptions applied to the secondary component of a cataclysmic variable, a best-fitting map of the surface of the secondary can be obtained.

## 2. Roche tomography in practice

To obtain a good map requires the best available spectral resolution and the highest possible signal-to-noise (S/N). Even so, maps produced using only a single line, applied successfully to map spots in relatively bright stars such as the RS CVn systems [2], reveal in the much fainter cataclysmic binaries only gross features, such as irradiation from the accretion flow. Fortunately, techniques have more recently been developed for single rapidly rotating stars [3] that enable the combination of many hundreds of lines using a technique called Least Squares Deconvolution (LSD), where an average line profile is calculated for each spectrum; this can increase the S/N ratio by a factor of 20 or more, and has revealed spots on several stars.

To obtain enough lines, a wide spectral range is needed as well as high resolution, making an echelle spectrograph the optimum choice. However, the continuum comes mostly from the other components in the system: the accretion disc and flow, and the white dwarf. These components vary in flux significantly with time, mostly irregularly with phase. To obtain a map consistent with the assumption of no intrinsic variations during the observations, these external continuum variations from one spectrum to the next must be removed. This requires either a comparison star on the slit (not usually practicable with an echelle) or simultaneous photometry, which is often not available at the same observing site.

### 2.1 What can be done if simultaneous photometry is not available?

To overcome the problem of obtaining simultaneous photometry, Hill et al. [4] have devised a scaling procedure (section 5.1 of their paper) that relies on the fact that Roche tomography uses only relative fluxes. An optimal subtraction method is used. Firstly, from the whole time series of spectra, a benchmark average (LSD) profile is chosen as being the closest to a Gaussian profile. This profile is then scaled and subtracted from its nearest neighbour in the time series, the optimal scaling factor being determined by minimising the residuals. This scaling factor is then applied to the neighbouring profile, and the process is repeated along the whole time series of LSD profiles, using the newly scaled profile as the comparison for the next step. This process removes the external flux variations in the continuum without affecting the *shape* of the profile, which is what determines the spot distribution.

### 2.2 Further practical details

The maximum entropy code requires a constraint if it is to produce a unique solution. This is implemented by the use of a regularisation statistic, that compares the data map with a 'default map' (for details see Section 3.3 of [5]; usually either the most uniform map consistent with the data, or the smoothest map consistent with the data, is chosen).

The final map takes account of limb darkening and gravity darkening as well as the Roche geometry. It will show any effects of irradiation from the other components in the system (white dwarf, disc, accretion flow), which will also show up as dark regions because the irradiation reduces the effective line flux both by increasing the overall continuum and, more importantly, by ionising (and thereby effectively removing) some fraction of the absorbing atoms.

### 3. CVs with LSD maps

In the early stages of developing Roche Tomography, five CVs had their secondaries mapped without the use of LSD, i.e. using just one strong absorption line. These were the novalike DW UMa, the dwarf nova IP Peg, and three polars: AM Her, HU Aqr and QQ Vul. These successfully showed the effects of irradiation, but no spots were detected ([5], [6]).

| Name | Type | Orbital period (h) | V magnitude | References |
|---|---|---|---|---|
| AE Aquarii | NL | 9.88 | 10.7 | [7], [8], [9], [4] |
| BV Centauri | DN, UG | 14.7 | 12.6 | [9], [10] |
| V426 Ophiuchi | DN, ZCam, IP? | 6.85 | 13.5 | [9], [11] |
| RU Pegasi | DN, UG | 8.99 | 12.5 | [12] |
| SS Cygni | DN, UG | 6.6 | 11.2 | |

Table 1: Systems with LSD maps published or being analysed

The use of LSD has enabled spot maps to be produced, and this has now resulted in published maps for four systems (Table 1). Spectra for a fifth system, SS Cygni (details in the table), are currently being analysed. A preliminary spectrum was presented at the 2013 Palermo meeting [13].

The four systems with published maps were all taken with different instruments at different times, but all show similar features: irradiation at the $L_1$ point plus a considerable number of spots, with one particularly large one at high latitude. Illustration 2 shows a comparison of the four maps, shown as they would be seen looking down on the north pole. The RU Peg map looks the smoothest because the data were taken at lower resolution.

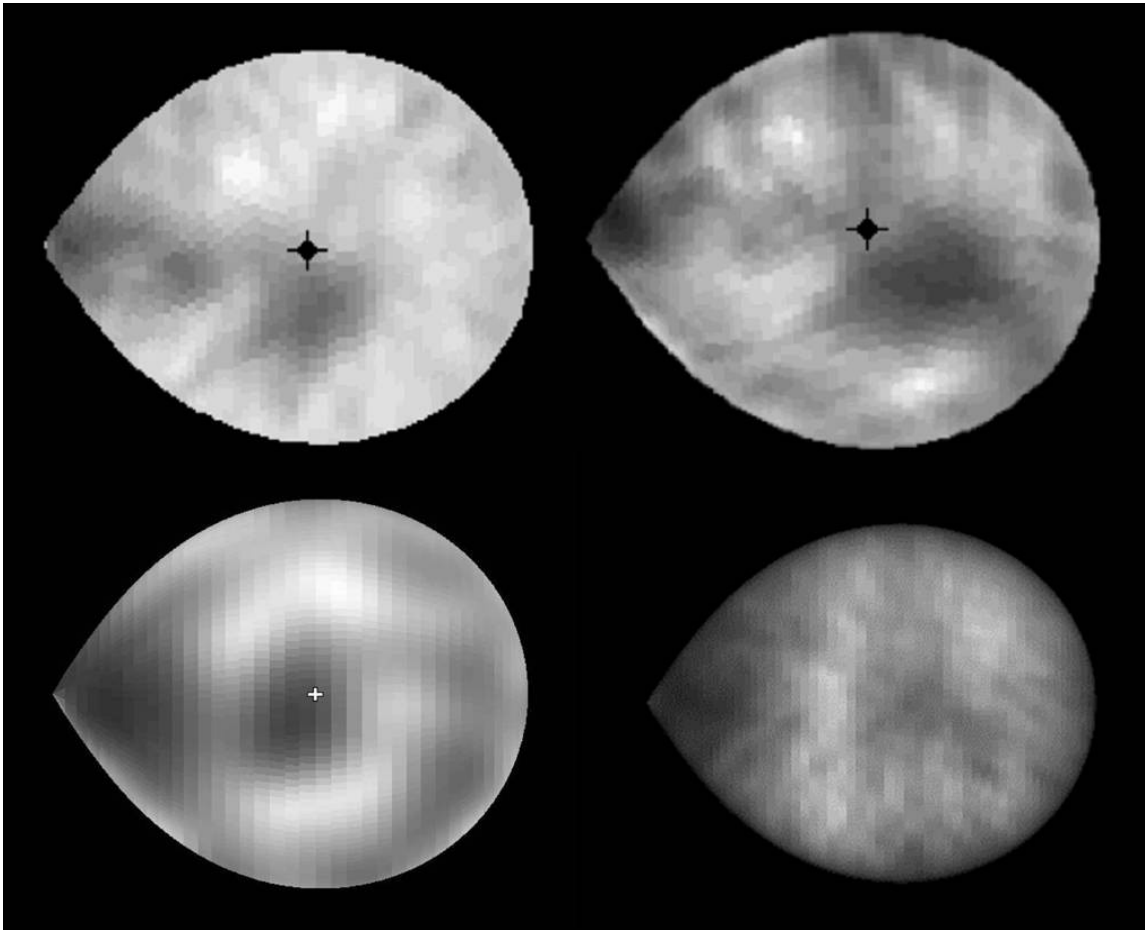

*Illustration 2: Comparison of maps for AE Aqr (top left), BV Cen (top right), RU Peg (bottom left) and V426 Oph (bottom right), seen from above the north pole.*

Maps of AE Aqr from a range of dates, and with a range of instruments show that a high-latitude spot is a constant feature, but that it moves in longitude, as shown in Illustration 3. The 2001 data were taken with the WHT and the UES, and the best-fitting inclination was found to be 66°. In 2004, 2005 and 2006 the data were taken with the Magellan telescope and the MIKE spectrograph, the best-fitting inclinations being, respectively, 59°, 59° and 60°. In 2008 and 2009 the data were taken with the VLT and UVES and yielded the even lower inclinations of 57°, 50° and 51°. This demonstrates that the inclination is poorly constrained by the maximum entropy technnique. The mass ratio *q* is in the range 0.64-0.68.

In 2009, the two datasets were taken about a week apart, allowing for the first time the determination of whether or not there is differential rotation [4].

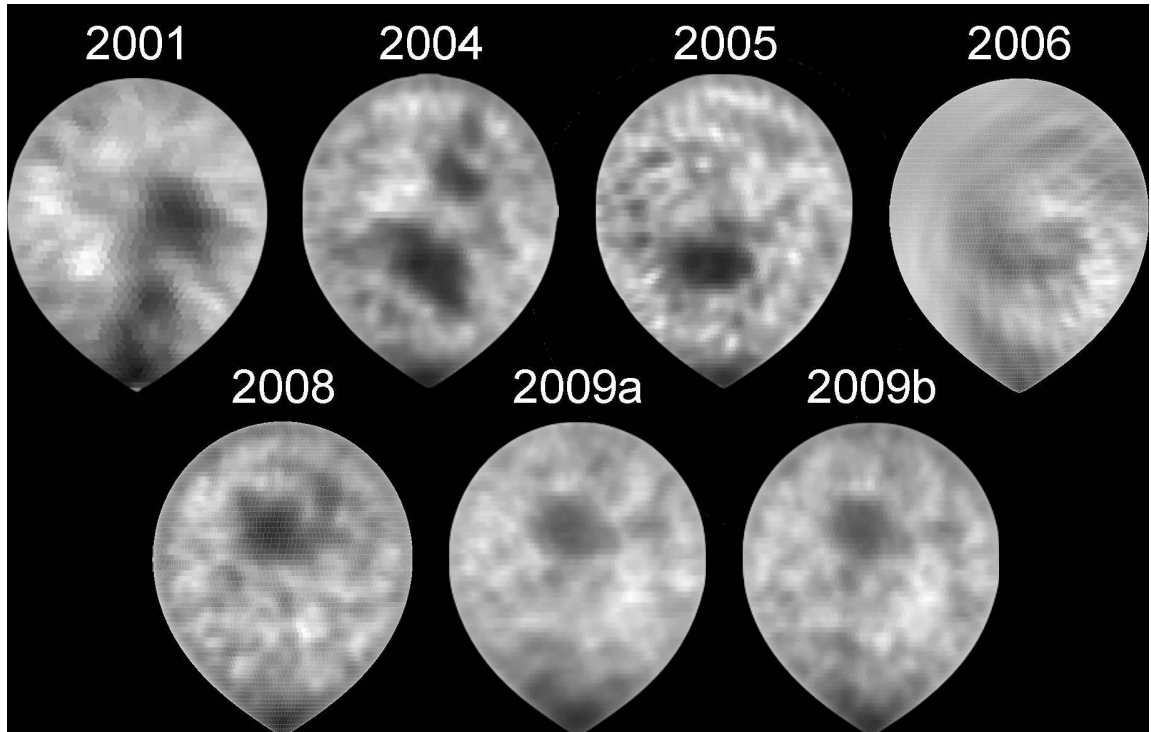

*Illustration 3: Maps of AE Aqr taken on a range of dates (see text for details). All show an overall spot distribution and a dominant high-latitude spot, at various longitudes.*

## Acknowledgements

I am grateful to the Royal Astronomical Society for partial travel and subsistence support for attending this meeting.

## References

[1] R.G.M. Rutten & V.S. Dhillon, *Roche tomography: imaging the stars in interacting binaries, A&A* **288** (773) 1994.

[2] S.S. Vogt & G.D. Penrod, *Doppler imaging of spotted stars: application to the RS Canum Venaticorum star HR 1099, PASP* **95** (565) 1983.

[3] A. Collier Cameron, *Spot mapping in cool stars,* in *Astrotomography,* eds H.M.J. Boffin, D. Steeghs & J. Cuypers, Springer 2001, p.183.

[4] C.A. Hill, C.A. Watson, T. Shahbaz, D. Steeghs & V.S. Dhillon, *Roche tomography of cataclysmic variables – VI. Differential rotation of AE Aqr – not tidally locked!, MNRAS* **444** (192) 2014.

[5] V.S. Dhillon & C.A. Watson, *Imaging the Secondary Stars in Cataclysmic Variables*, in *Astrotomography,* eds H.M.J. Boffin, D. Steeghs & J. Cuypers, Springer 2001, p.94.

[6] C.A. Watson, V.S. Dhillon, R.G.M. Rutten & A.D. Schwope, *Roche tomography of cataclysmic variables – II. Images of the secondary stars in AM Her, QQ Vul, IP Peg and HU Aqr, MNRAS* **341** (129) 2003

[7] C.A. Watson & V.S. Dhillon, *Roche tomography of the secondary stars in CVs, Astron. Nachr.* **325** (189) 2004.

[8] C.A. Watson, V.S. Dhillon & T. Shahbaz, *Roche tomography of cataclysmic variables – III. Starspots on AE Aqr, MNRAS* **368** (637) 2006.

[9] C.A. Watson, D. Steeghs, V.S Dhillon & T. Shahbaz, *Imaging the cool stars in the interacting binaries AE Aqr, BV Cen and V426 Oph*, *Astron. Nachr.* **328** (813) 2007.

[10] C.A. Watson, D. Steeghs, T. Shahbaz & V.S. Dhillon, *Roche tomography of cataclysmic variables – IV. Starspots and slingshot prominences on BV Cen, MNRAS* **382** (1105) 2007.

[11] C.A Hill, *unpublished MSci project, Queen's University Belfast,* 2012; paper in preparation.

[12] A. Dunford, C.A. Watson & R.C. Smith, *Roche tomography of cataclysmic variables – V. A high-latitude starspot on RU Pegasi, MNRAS* **422** (3444) 2012.

[13] R.C. Smith, J. Echevarria, J. Venancio Hernandez, & P. Szkody, *SS Cygni revisited*, in *The Golden Age of Cataclysmic Variables and Related Objects 2*, eds F. Giovannelli & L. Sabau-Graziati, *Acta Polytechnica CTU Proceedings,* **2** (148) 2015.

**DISCUSSION**

**DAVID BUCKLEY:** Some years ago, people working on RS Cvn data expressed some concerns that polar spots inferred from similar studies may be artefacts of the deconvolution process. Also, from theory they were claiming that polar spots were unlikely.

**ROBERT SMITH:** CV secondaries are rapidly rotating objects, with rotation periods equal to the orbital period, and typical surface rotation speeds of 100 km/s. Observational studies of single rapidly rotating stars also show high-latitude spots. In addition, dynamo theory suggests that spot latitudes increase as rotation speed increases. So I think polar spots are to be expected, and the observed features are real.

**MICHELE MONTGOMERY:** 1. What is the percentage spot coverage on AE Aqr? 2. Are these spots from magnetic activity, where they appear and disappear on a roughly 2-week timescale, and in different locations? 3. Is there a period of drift, since the spot seems to drift in longitude? 4. Why are the spots nearly always on the trailing hemisphere? 5. How do we know the intensity of the spot? 6. What is the minimum angular size that can be be resolved?

**ROBERT SMITH:** 1. According to Hill et al. [4], the spot coverage for AE Aqr was between 15% and 17% of the northern hemisphere in August and September 2009, consistent with earlier measurements of 18%. 2. We believe that the spots arise from magnetic activity, but the evidence suggests that they are larger and longer-lived than sunspots. 3. There are not enough data, even for AE Aqr, to follow the drifts in longitude in detail. 4. For AE Aqr, the spots are not always on the trailing hemisphere, so perhaps it is just a coincidence that the other three CVs happened to have spots on the trailing hemisphere when observed – small number statistics! 5. The intensity of the spots comes from the relative flux of the bumps in the line profile, so good flux calibration is essential. 6. The minimum angular resolution that can be resolved in longitude depends on the phase resolution of the spectra. For data taken with a large telescope and short exposures, it could be as small as a few degrees.